\documentclass[aps,prd,twocolumn,superscriptaddress]{revtex4}
\usepackage{epsfig,epsf}
\usepackage{amsmath}
\usepackage{amsthm}
\usepackage{amsfonts}
\usepackage{amssymb}
\usepackage{dsfont}
\usepackage{multirow}
\usepackage{appendix}
\usepackage{slashed}
\usepackage[active]{srcltx}
\usepackage{psfrag}

\begin{document}

\title{Semileptonic and nonleptonic decays of the axial-vector tetraquark $%
T_{bb;\overline{u} \overline{d}}^{-}$}
\date{\today}
\author{S.~S.~Agaev}
\affiliation{Institute for Physical Problems, Baku State University, Az--1148 Baku,
Azerbaijan}
\author{K.~Azizi}
\affiliation{Department of Physics, University of Tehran, North Karegar Avenue, Tehran
14395-547, Iran}
\affiliation{Department of Physics, Do\v{g}u\c{s} University, Acibadem-Kadik\"{o}y, 34722
Istanbul, Turkey}
\author{B.~Barsbay}
\affiliation{Department of Physics, Do\v{g}u\c{s} University, Acibadem-Kadik\"{o}y, 34722
Istanbul, Turkey}
\affiliation{Department of Physics, Kocaeli University, 41380 Izmit, Turkey}
\author{H.~Sundu}
\affiliation{Department of Physics, Kocaeli University, 41380 Izmit, Turkey}

\begin{abstract}
The semileptonic and nonleptonic decays of the double-beauty axial-vector
tetraquark $T_{bb;\overline{u} \overline{d}}^{-}$ to a state $\widetilde{T}_{bc;%
\overline{u}\overline{d}}^{0}$ (hereafter $T_{bb}^{-}$ and $\widetilde{T}%
_{bc}^{0}$, respectively) are investigated in the context of the QCD sum
rule method. The final-state tetraquark $\widetilde{T}_{bc}^{0}$ is treated
as an axial-vector particle built of a heavy axial-vector diquark $%
b^{T}\gamma _{\mu }Cc$ and light scalar antidiquark $\overline{u}C\gamma_{5}%
\overline{d}^{T}$. Its spectroscopic parameters are calculated using the
two-points sum rules by taking into account contributions of quark, gluon
and mixed condensates up to dimension $10$. We study the dominant
semileptonic $T_{bb}^{-} \to \widetilde{T}_{bc}^{0}l\overline{\nu }_{l}$ and
nonleptonic decays $T_{bb}^{-} \to \widetilde{T}_{bc}^{0}M$, where $M$ is
one of the pseudoscalar mesons $\pi^{-}, K^{-}, D^{-}$ and $D_s^{-}$. The
partial widths of these processes are computed in terms of weak form factors
$G_{i}(q^2),\ i=1,2,3,4$, extracted by employing the QCD three-point sum
rule approach. Predictions obtained for partial widths of considered decays
are used to improve accuracy of theoretical predictions for full width and
lifetime of the tetraquark $T_{bb}^{-}$, which are important for
experimental exploration of this exotic meson.
\end{abstract}

\maketitle


\section{Introduction}

\label{sec:Int}
The four-quark exotic mesons containing a few heavy $Q=b$, $c$ quarks are
particles, investigation of which attracted interest of scientists more than
thirty years ago \cite%
{Iwasaki:1977qw,Chao:1980dv,Ader:1981db,Lipkin:1986dw,Zouzou:1986qh,Heller:1985cb,Carlson:1987hh}%
. Among these particles most interesting are states composed of heavy
diquarks $QQ^{\prime }$ and light $\overline{q}\overline{q}^{\prime }$
antidiquarks, because they are real candidates to stable exotic mesons.
First qualitative results concerning a stability of the compounds $%
QQ^{\prime }\overline{q}\overline{q}^{\prime }$ against strong decays were
obtained already in Refs. \cite%
{Ader:1981db,Lipkin:1986dw,Zouzou:1986qh,Carlson:1987hh}. Exotic mesons $%
QQ^{\prime }\overline{q}\overline{q}^{\prime }$ may decay through strong
interaction to mesons $Q\overline{q}$ and $Q^{\prime }\overline{q}^{\prime }$
or to $Q\overline{q}^{\prime }$ and $Q^{\prime }\overline{q}$ if the mass of
the master particle exceeds masses of final-state conventional mesons. It
was demonstrated that such four-quark states would be stable provided that
the ratio $m_{Q}/m_{q}$ is sufficiently large. A well known particle from
this range is the axial-vector tetraquark $T_{bb}^{-}$: Calculations carried
out in the context of different models proved that its mass is below the $B%
\overline{B}^{\ast }$ threshold, and $T_{bb}^{-}$ is the particle stable
against strong decays \cite{Carlson:1987hh,Navarra:2007yw}.

During the last decade properties of tetraquarks composed of heavy $bb$ and $%
bc$ diquarks were investigated in numerous articles using various methods
(see, for example Refs.\ \cite%
{Du:2012wp,Chen:2013aba,Francis:2018jyb,Caramees:2018oue} and references
therein). Recent analysis of the heavy-light particles $QQ^{\prime }%
\overline{q}\overline{q}^{\prime }$ confirmed stable nature of the
tetraquark $T_{bb}^{-}$ \cite{Karliner:2017qjm,Eichten:2017ffp,Agaev:2018khe}%
. In our work \cite{Agaev:2018khe} the axial-vector particle $T_{bb}^{-}$
was studied by means of the QCD sum rule method. In accordance with our
result, the mass of $T_{bb}^{-}$ is equal to $m=(10035\pm 260)~\mathrm{MeV}$
which is below the $B^{-}\overline{B}^{\ast 0}$ and $B^{-}\overline{B}%
^{0}\gamma $ thresholds. In other words, this particle is stable against the
strong and radiative decays. Hence, it dissociates to conventional mesons
via weak processes considered also in Ref.\ \cite{Agaev:2018khe}. We
explored the semileptonic decays $T_{bb}^{-}$ $\rightarrow Z_{bc}^{0}l%
\overline{\nu }_{l}$, where the final-state tetraquark $Z_{bc}^{0}=[bc][%
\overline{u}\overline{d}]$ was treated as a scalar particle. By computing
partial widths of these decays, we estimated the width $\Gamma $ and mean
lifetime $\tau $ of the axial-vector tetraquark $T_{bb}^{-}$. Predictions
obtained for these parameters $\Gamma =(7.17\pm 1.23)\times 10^{-8}~\mathrm{%
MeV}$ and $\tau =9.18_{-1.34}^{+1.90}~\mathrm{fs}$ may be useful for
experimental investigation of double-heavy exotic mesons. Problems connected
with calculation of parameters and weak decay channels of $\ T_{bb}^{-}$
were addressed in Ref.\ \cite{Hernandez:2019eox}, as well.

Investigations showed that not only the tetraquark $T_{bb}^{-}$, but also
other double-beauty states may be strong and electromagnetic interactions
stable particles. Thus, the scalar counterparts of $T_{bb}^{-}$, i.e., the
tetraquark $T_{b:\overline{d}}^{-}$, the scalar and axial-vector four-quark
mesons $T_{bb;\overline{u}\overline{s}}^{-}$ are stable against strong and
radiative decays. The spectroscopic parameters, widths and lifetimes of
these exotic mesons were computed in Refs.\ \cite%
{Agaev:2020dba,Agaev:2019lwh,Agaev:2020zag}.

As we have mentioned above, tetraquarks containing a diquark $bc$ are also
interesting object for studies, because some of them may be stable
particles. Thus, the mass of the scalar exotic meson $Z_{bc}^{0}$ equals to $%
m_{Z}=(6660\pm 150)~\mathrm{MeV}$, which is below thresholds for strong and
radiative decays of $Z_{bc}^{0}$ to conventional mesons \cite{Agaev:2018khe}%
. As a result, $Z_{bc}^{0}$ is the strong- and electromagnetic-interaction
stable compound weak decays of which were studied in Ref.\ \cite%
{Sundu:2019feu}. Predictions for width and lifetime of $Z_{bc}^{0}$ obtained
there provide valuable information on features of this particle. The
spectroscopic parameters and possible strong and weak decay channels of the
axial-vector tetraquark $T_{bc}^{0}=bc\overline{u}\overline{d}$ was studied,
as well \cite{Agaev:2019kkz}.

In the present article, we extend our analysis of the tetraquark $T_{bb}^{-}$
by considering its new weak decay channels $T_{bb}^{-} \rightarrow
\widetilde{T}_{bc}^{0}l\overline{\nu }_{l}$ and $T_{bb}^{-} \rightarrow
\widetilde{T}_{bc}^{0}M$, where $\widetilde{T}_{bc}^{0}$ is an axial-vector
state. This investigation will allow us to improve estimates for the full
width and lifetime of $T_{bb}^{-}$.

We treat $\widetilde{T}_{bc}^{0}$ as a tetraquark composed of
color-antitriplet heavy axial-vector diquark $bc$ and light color-triplet
scalar antidiquark $\overline{u}\overline{d}$. It is worth noting that quark
contents and quantum numbers of the tetraquarks $T_{bc}^{0}$ and $\widetilde{%
T}_{bc}^{0}$ are the same, and both of them have the antisymmetric color
structure $[\overline{\mathbf{3}}_{c}]_{bc}\otimes \lbrack \mathbf{3}_{c}]_{%
\overline{u}\overline{d}}$. But $T_{bc}^{0}$ and $\widetilde{T}_{bc}^{0}$
differ from each another due to their internal organizations. In fact, the
heavy diquark in $T_{bc}^{0}$ is a scalar state, whereas the tetraquark $%
\widetilde{T}_{bc}^{0}$ is made of an axial-vector heavy diquark $bc$. The
reason for such choice of the final-state tetraquark $\widetilde{T}_{bc}^{0}$
will be explained later. The axial-vector tetraquarks  $[bc][\overline{q}\overline{q}^{\prime }]$ with definite isospins  can be modeled using a structure 
$\widetilde{T}_{bc}^{0}$. But in the limit $ m_u=m_d=0 $, adopted in the present study, these states are degenerate and their parameters do not differ from the ones of $\widetilde{T}_{bc}^{0}$. This conclusion is valid for the tetraquarks $T_{bb}^{-}$ and  $T_{bc}^{0}$ as well. 

To compute partial widths of the initial $T_{bb}^{-}$ particle's weak
decays, apart from its mass and current coupling, one needs also spectral
parameters of the tetraquark $\widetilde{T}_{bc}^{0}$. The mass $\widetilde{m%
}$ and coupling $\widetilde{f}$ of the state $\widetilde{T}_{bc}^{0}$ are
calculated using the QCD sum rule method \cite{Shifman:1978bx,Shifman:1978by}%
, which is a powerful tool to calculate parameters of conventional mesons
and baryons. It can be successfully applied to analyze multiquark hadrons,
as well \cite{Agaev:2016dev,Albuquerque:2018jkn}. We calculate the mass and
current coupling of the tetraquark $\widetilde{T}_{bc}^{0}$ using relevant
interpolating current by taking into account various quark, gluon, and mixed
condensates up to dimension $10$. Spectral parameters of $\widetilde{T}%
_{bc}^{0}$ extracted from such analysis are also of particular interest to
explore the family of tetraquarks $bc\overline{q}\overline{q}^{\prime }$.

There are different weak decays of $T_{bb}^{-}$, but dominant ones are
processes triggered by a subprocess $b\rightarrow W^{-}c$ responsible for
transformation of $T_{bb}^{-}$ to the final axial-vector state $\widetilde{T}%
_{bc}^{0}$. In semileptonic decays $T_{bb}^{-}\rightarrow \widetilde{T}%
_{bc}^{0}l\overline{\nu }_{l}$ the tetraquark $\widetilde{T}_{bc}^{0}$ is
accompanied by a lepton pair $l\overline{\nu }_{l}$, whereas in nonleptonic
processes $T_{bb}^{-} \rightarrow \widetilde{T}_{bc}^{0}M$ there is an
additional ordinary meson $M$ in the final phase of the process. We consider
decays in which $M$ is one of the conventional pseudoscalar mesons $\pi ^{-}
$, $K^{-}$, $D^{-}$ and $D_{s}^{-}$. To evaluate partial widths of weak
decays one has to determine form factors $G_{i}(q^{2}),\ i=1,2,3,4$ which
govern weak transitions: They enter to differential rate $d\Gamma /dq^{2}$
of semileptonic and partial width of nonleptonic processes. To this end, we
employ the QCD three-point sum rule approach, and extract $G_{i}(q^{2})$ at $%
q^{2\text{ }}$ accessible for sum rule calculations. As usual, these $q^{2}$
do not cover a full region $m_{l}^{2}\leq q^{2}\leq (m-\widetilde{m})^{2}$
necessary to integrate the differential rates $d\Gamma /dq^{2}$ of
semileptonic decays. Therefore, one has to introduce model functions $%
\mathcal{G}_{i}(q^{2})$ that coincide with the sum rule predictions when
they are accessible, and can be easily extrapolated to all $q^{2}$: Usage of
$\mathcal{G}_{i}(q^{2})$ in calculations solves these technical problems.

This article is structured in the following way: In Sec.\ \ref{sec:Mass}, we
evaluate the mass and current coupling of the tetraquark $\widetilde{T}%
_{bc}^{0}$ by employing the QCD two-point sum rule method. Calculations of
the weak form factors $G_{i}(q^{2})$ in the framework of the three-point sum
rule approach are performed in section \ref{sec:WFF}. Here, we determine the
model functions $\mathcal{G}_{i}(q^{2})$ and also find partial widths of the
semileptonic decays $T_{bb}^{-} \rightarrow \widetilde{T}_{bc}^{0}l\overline{%
\nu }_{l}$. Section \ref{sec:NLDecays} is devoted to analysis of the
nonleptonic weak transformations of the tetraquark $T_{bb}^{-}$. In section %
\ref{sec:Disc} we calculate the full width and lifetime of $T_{bb}^{-}$, and
discuss obtained results. This section contains also our concluding notes.


\section{Mass and current coupling of the axial-vector tetraquark $%
\widetilde{T}_{bc}^{0}$}

\label{sec:Mass}

The mass $\widetilde{m}$, and coupling $\widetilde{f}$ of the tetraquark $%
\widetilde{T}_{bc}^{0}$ are important parameters of the problem under
consideration: they are required to find partial widths of the weak
processes $T_{bb}^{-}$ $\rightarrow \widetilde{T}_{bc}^{0}l\overline{\nu }%
_{l}$ and $T_{bb}^{-}$ $\rightarrow \widetilde{T}_{bc}^{0}M$. Besides, the
axial-vector tetraquark $\widetilde{T}_{bc}^{0}$, as its partner state $%
T_{bc}^{0}$, may be strong- and/or electromagnetic-interaction stable
particle, which is interesting in itself.

The sum rules to extract spectroscopic parameters of $\widetilde{T}_{bc}^{0}$
can be derived from analysis of the two-point correlation function $\Pi
_{\mu \nu }(p)$ given by the expression
\begin{equation}
\Pi _{\mu \nu }(p)=i\int d^{4}xe^{ipx}\langle 0|\mathcal{T}\{\widetilde{J}%
_{\mu }(x)\widetilde{J}_{\nu }^{\dag }(0)\}|0\rangle,  \label{eq:CF1}
\end{equation}%
where $\widetilde{J}_{\mu }(x)$ is the interpolating current to the
axial-vector tetraquark $\widetilde{T}_{bc}^{0}$. The structure of this
current is determined, in some respects, by the organization of the initial
particle $T_{bb}^{-}$. It is instructive to consider the structure and
interpolating current $J_{\mu }(x)$ of the teraquark $T_{bb}^{-}$
\begin{equation}
J_{\mu }(x)=b_{a}^{T}(x)\gamma _{\mu }Cb_{b}(x)\overline{u}_{a}(x)C\gamma
_{5}\overline{d}_{b}^{T}(x),  \label{eq:Curr1}
\end{equation}%
where $a$ and $b$ are color indices and $C$ is charge-conjugation operator.
The current $J_{\mu }(x)$ was used in Ref.\ \cite{Agaev:2018khe} to study
the exotic meson $T_{bb}^{-}$. As is seen, $T_{bb}^{-}$ is built of the
axial-vector diquark $b^{T}\gamma _{\mu }Cb$ and light scalar antidiquark $%
\overline{u}C\gamma _{5}\overline{d}^{T}$, and the current $J_{\mu }$
belongs to the $[\overline{\mathbf{3}}_{c}]_{bb}\otimes \lbrack \mathbf{3}%
_{c}]_{\overline{u}\overline{d}}$ representation of the color group $%
SU_{c}(3)$.

Weak decays of $T_{bb}^{-}$ run through transition of $b$-quark $%
b\rightarrow c$, therefore expected organization of the final diquark field
is $b_{a}^{T}\gamma _{\mu }Cc_{b}$, whereas the antidiquark field preserves
its quark content and scalar nature. Then, at the final state we get the
axial-vector tetraquark, which is described by the current
\begin{eqnarray}
\widetilde{J}_{\mu }(x) &=&b_{a}^{T}(x)\gamma _{\mu }Cc_{b}(x)\left[
\overline{u}_{a}(x)C\gamma _{5}\overline{d}_{b}^{T}(x)\right.  \notag \\
&&\left. -\overline{u}_{b}(x)\gamma _{\mu }C\gamma _{5}\overline{d}%
_{a}^{T}(x)\right].  \label{eq:Curr2}
\end{eqnarray}%
The current $\widetilde{J}_{\mu }$ is antisymmetric in color indices and has
color-triplet structure $[\overline{\mathbf{3}}_{c}]_{bc}\otimes \lbrack
\mathbf{3}_{c}]_{\overline{u}\overline{d}}$. It is known that scalar and
axial vector triplet diquarks are most favorable two-quark states to
construct tetraquarks with $J^{\mathrm{P}}=1^{+}$ \cite{Jaffe:2004ph}. The
current $\widetilde{J}_{\mu }$ corresponds to lower lying tetraquark with
structure $\gamma _{\mu }C\otimes C\gamma _{5}$ and spin-parity $J^{\mathrm{P%
}}=1^{+}$. Of course, there is an alternative choice for $\widetilde{J}_{\mu
}$ composed of a heavy scalar diquark and an axial-vector light antidiquark.
Properties of such state $T_{bc}^{0}$ with a composition $C\gamma
_{5}\otimes \gamma _{\mu }C$, its weak and strong decays were investigated
in Ref. \cite{Agaev:2019kkz}. To model $\widetilde{T}_{bc}^{0}$ we choose $%
\widetilde{J}_{\mu }$ given by Eq.\ (\ref{eq:Curr2}) as a current stemming
naturally from organization of the master particle $T_{bb}^{-}$.

To find the sum rules for the mass $\widetilde{m}$ and coupling $\widetilde{f%
}$ of the tetraquark $\widetilde{T}_{bc}^{0}$, we write down the correlation
function $\Pi _{\mu \nu }^{\mathrm{Phys}}(p)$ using physical parameters of $%
\widetilde{T}_{bc}^{0}$. We treat $\widetilde{T}_{bc}^{0}$ as a ground-state
particle, and separate its contribution to $\Pi _{\mu \nu }^{\mathrm{Phys}%
}(p)$ from other terms
\begin{equation}
\Pi _{\mu \nu }^{\mathrm{Phys}}(p)=\frac{\langle 0|\widetilde{J}_{\mu }|%
\widetilde{T}_{bc}^{0}(p)\rangle \langle \widetilde{T}_{bc}^{0}(p)|%
\widetilde{J}_{\nu }^{\dagger }|0\rangle }{\widetilde{m}^{2}-p^{2}}+\cdots.
\label{eq:CF2}
\end{equation}%
Effects of higher resonances and continuum states are denoted in $\Pi _{\mu
\nu }^{\mathrm{Phys}}(p)$ by dots. The expression \ (\ref{eq:CF2}) is
derived by saturating the correlation function with a complete set of $%
J^{P}=1^{+}$ states with required quark content and performing integration
in $\Pi _{\mu \nu }(p)$ over $x$.

The correlator $\Pi _{\mu \nu }^{\mathrm{Phys}}(p)$ can be simplified by
introducing the matrix element $\langle 0|\widetilde{J}_{\mu }|\widetilde{T}%
_{bc}^{0}(p)\rangle $
\begin{equation}
\langle 0|\widetilde{J}_{\mu }|\widetilde{T}_{bc}^{0}(p)\rangle =\widetilde{m%
}\widetilde{f}\widetilde{\epsilon }_{\mu },  \label{eq:MElem1}
\end{equation}%
where $\widetilde{\epsilon }_{\mu }$ is the polarization vector of the
tetraquark $\widetilde{T}_{bc}^{0}$. In terms of the mass $\widetilde{m}$
and coupling $\widetilde{f}$ the function $\Pi _{\mu \nu }^{\mathrm{Phys}%
}(p) $ takes the form
\begin{equation}
\Pi _{\mu \nu }^{\mathrm{Phys}}(p)=\frac{\widetilde{m}^{2}\widetilde{f}^{2}}{%
\widetilde{m}^{2}-p^{2}}\left( -g_{\mu \nu }+\frac{p_{\mu }p_{\nu }}{%
\widetilde{m}^{2}}\right) +\cdots.  \label{eq:CorM}
\end{equation}

The sum rules require computation of $\Pi _{\mu \nu }(p)$ in terms of quark
propagators, as well. To this end, one needs to substitute $\widetilde{J}%
_{\mu }(x)$ into the correlation function (\ref{eq:CF1}) and contract
relevant light and heavy quark fields. These manipulations yields
\begin{eqnarray}
\Pi _{\mu \nu }^{\mathrm{OPE}}(p) &=&i\int d^{4}xe^{ipx}\mathrm{Tr}\left[
\gamma _{\mu }\widetilde{S}_{b}^{aa^{\prime }}(x)\gamma _{\nu
}S_{c}^{bb^{\prime }}(x)\right]  \notag \\
&&\times \left\{ \mathrm{Tr}\left[ \gamma _{5}\widetilde{S}_{d}^{a^{\prime
}b}(-x)\gamma _{5}S_{u}^{b^{\prime }a}(-x)\right] \right.  \notag \\
&&-\mathrm{Tr}\left[ \gamma _{5}\widetilde{S}_{d}^{b^{\prime }b}(-x)\gamma
_{5}S_{u}^{a^{\prime }a}(-x)\right]  \notag \\
&&+\mathrm{Tr}\left[ \gamma _{5}\widetilde{S}_{d}^{b^{\prime }a}(-x)\gamma
_{5}S_{u}^{a^{\prime }b}(-x)\right]  \notag \\
&&\left. -\mathrm{Tr}\left[ \gamma _{5}\widetilde{S}_{d}^{a^{\prime
}a}(-x)\gamma _{5}S_{u}^{b^{\prime }b}(-x)\right] \right\},
\label{eq:CF3vector}
\end{eqnarray}%
where $S_{Q}^{ab}(x)$ and $S_{q}^{ab}(x)$ are the heavy $Q=b(c)$ and light $%
q=d(u)$ quark propagators, respectively. Here, we have introduced also the
notation
\begin{equation}
\widetilde{S}_{Q(q)}(x)=CS_{Q(q)}^{T}(x)C.  \label{eq:Prop}
\end{equation}%
In the present work, we use the light quark propagator given by the
expression \cite{Agaev:2020zad}
\begin{eqnarray}
&&S_{q}^{ab}(x)=i\frac{\slashed x\delta _{ab}}{2\pi ^{2}x^{4}}-\frac{%
m_{q}\delta _{ab}}{4\pi ^{2}x^{2}}-\frac{\langle \overline{q}q\rangle }{12}%
\left( 1-i\frac{m_{q}}{4}\slashed x\right) \delta _{ab}  \notag \\
&&-\frac{x^{2}}{192}\langle \overline{q}g_{s}\sigma Gq\rangle \left( 1-i%
\frac{m_{q}}{6}\slashed x\right) \delta _{ab}-\frac{\slashed xx^{2}g_{s}^{2}%
}{7776}\langle \overline{q}q\rangle ^{2}\delta _{ab}  \notag \\
&&-\frac{ig_{s}G_{ab}^{\mu \nu }}{32\pi ^{2}x^{2}}\left[ \slashed x\sigma
_{\mu \nu }+\sigma _{\mu \nu }\slashed x\right] -\frac{x^{4}\langle
\overline{q}q\rangle \langle g_{s}^{2}G^{2}\rangle }{27648}\delta
_{ab}+\cdots.  \notag \\
&&  \label{eq:qProp}
\end{eqnarray}%
The propagator of the heavy quarks $Q$ is determined by the formula
\begin{eqnarray}
&&S_{Q}^{ab}(x)=i\int \frac{d^{4}k}{(2\pi )^{4}}e^{-ikx}\Bigg \{\frac{\delta
_{ab}\left( {\slashed k}+m_{Q}\right) }{k^{2}-m_{Q}^{2}}  \notag \\
&&-\frac{g_{s}G_{ab}^{\alpha \beta }}{4}\frac{\sigma _{\alpha \beta }\left( {%
\slashed k}+m_{Q}\right) +\left( {\slashed k}+m_{Q}\right) \sigma _{\alpha
\beta }}{(k^{2}-m_{Q}^{2})^{2}}  \notag \\
&&+\frac{g_{s}^{2}G^{2}}{12}\delta _{ab}m_{Q}\frac{k^{2}+m_{Q}{\slashed k}}{%
(k^{2}-m_{Q}^{2})^{4}}+\frac{g_{s}^{3}G^{3}}{48}\delta _{ab}\frac{\left( {%
\slashed k}+m_{Q}\right) }{(k^{2}-m_{Q}^{2})^{6}}  \notag \\
&&\times \left[ {\slashed k}\left( k^{2}-3m_{Q}^{2}\right) +2m_{Q}\left(
2k^{2}-m_{Q}^{2}\right) \right] \left( {\slashed k}+m_{Q}\right) +\cdots %
\Bigg \}.  \notag \\
&&  \label{eq:QProp}
\end{eqnarray}%
In expressions (\ref{eq:qProp}) and (\ref{eq:QProp})
\begin{eqnarray}
G_{ab}^{\alpha \beta } &=&G_{A}^{\alpha \beta
}t_{ab}^{A},\,\,~~G^{2}=G_{\alpha \beta }^{A}G_{\alpha \beta }^{A},  \notag
\\
G^{3} &=&\,\,f^{ABC}G_{\mu \nu }^{A}G_{\nu \delta }^{B}G_{\delta \mu }^{C},
\end{eqnarray}%
where $a,\,b=1,2,3$ are color indices and $A,B,C=1,\,2\,\cdots 8$. Here $%
t^{A}=\lambda ^{A}/2$, where $\lambda ^{A}$ are the Gell-Mann matrices. The
gluon field strength tensor is fixed at $x=0$, i. e., $G_{\alpha \beta
}^{A}\equiv G_{\alpha \beta }^{A}(0)$.

To proceed one should choose invariant amplitudes corresponding to the same
Lorentz structures from both $\Pi _{\mu \nu }^{\mathrm{Phys}}(p)$ and $\Pi
_{\mu \nu }^{\mathrm{OPE}}(p)$. There are two Lorentz structures
proportional to $g_{\mu \nu }$ and $p_{\mu }p_{\nu }$ in these correlation
functions. Because invariant amplitudes $\Pi ^{\mathrm{Phys}}(p^{2})$ and $%
\Pi ^{\mathrm{OPE}}(p^{2})$ corresponding to terms $\sim g_{\mu \nu }$ do
not contain contributions of scalar particles, we work with these functions.
The sum rules for $\widetilde{m}$ and $\widetilde{f}$ can be obtained by
equating these invariant amplitudes and performing standard prescriptions of
the sum rule method. As the first step, one applies the Borel transformation
to both sides of obtained equality, which is necessary to suppress
contributions due to higher resonances and continuum states. Afterwards,
these contributions should be subtracted from the physical side of this
equality by employing the hypothesis on quark-hadron duality. After these
manipulations, a final expression becomes a function of the Borel $M^{2}$
and continuum threshold $s_{0}$ parameters. The second expression required
to derive sum rules for $\widetilde{m}$ and $\widetilde{f}$ can be obtained
from the first equality by acting on it by the operator $d/d(-1/M^{2})$. As
a result, for $\widetilde{m}$ and $\widetilde{f}$, we get the sum rules
\begin{equation}
\widetilde{m}^{2}=\frac{\Pi^{\prime } (M^{2},s_{0})}{\Pi (M^{2},s_{0})},
\label{eq:Mass}
\end{equation}%
and
\begin{equation}
\widetilde{f}^{2}=\frac{e^{\widetilde{m}^{2}/M^{2}}}{\widetilde{m}^{2}}\Pi
(M^{2},s_{0}).  \label{eq:Coupling}
\end{equation}%
In expressions above $\Pi (M^{2},s_{0})$ is the Borel transformed and
continuum subtracted invariant amplitude $\Pi ^{\mathrm{OPE}}(p^{2})$, and $%
\Pi ^{\prime }(M^{2},s_{0})=d/d(-1/M^{2})\Pi (M^{2},s_{0})$.

The function $\Pi (M^{2},s_{0})$ has the following form%
\begin{equation}
\Pi (M^{2},s_{0})=\int_{\mathcal{M}^{2}}^{s_{0}}ds\rho ^{\mathrm{OPE}%
}(s)e^{-s/M^{2}}+\Pi (M^{2}),  \label{eq:InvAmp}
\end{equation}%
where $\mathcal{M}=m_{b}+m_{c}$. Here, $\rho ^{\mathrm{OPE}}(s)$ is the
two-point spectral density computed as an imaginary part of the correlation
function. The second term in Eq.\ (\ref{eq:InvAmp}) contains nonperturbative
contributions computed directly from $\Pi_{\mu\nu} ^{\mathrm{OPE}}(p)$. In the
present work, we calculate $\Pi (M^{2},s_{0})$ by taking into account
nonperturbative terms up to dimension $10$. The explicit expression of the
function $\Pi (M^{2},s_{0})$ is rather lengthy, therefore we do not provide
it here.

The obtained sum rules contain numerous input parameters, which have to be
specified in order to carry out computations. The vacuum condensates and
masses of $b$, and $c$ quarks are universal parameters and do not depend on
the problem under analysis: Their values are listed below
\begin{eqnarray}
&&\langle \overline{q}q\rangle =-(0.24\pm 0.01)^{3}~\mathrm{GeV}^{3},\
\notag \\
&&\langle \overline{q}g_{s}\sigma Gq\rangle =m_{0}^{2}\langle \overline{q}%
q\rangle ,\ m_{0}^{2}=(0.8\pm 0.1)~\mathrm{GeV}^{2},  \notag \\
&&\langle \frac{\alpha _{s}G^{2}}{\pi }\rangle =(0.012\pm 0.004)~\mathrm{GeV}%
^{4},  \notag \\
&&\langle g_{s}^{3}G^{3}\rangle =(0.57\pm 0.29)~\mathrm{GeV}^{6},\   \notag
\\
&&m_{c}=1.27\pm 0.2~\mathrm{GeV},~m_{b}=4.18_{-0.02}^{+0.03}~\mathrm{GeV}.
\label{eq:Parameters}
\end{eqnarray}%
The mass and coupling of the tetraquark $\widetilde{T}_{bc}^{0}$ depend on
the auxiliary parameters $M^{2}$ and $s_{0}$, and their correct choice is
one of important problems of our studies. We fix the upper allowed value of
the Borel parameter from a restriction $\mathrm{PC}>0.2$, where $\mathrm{PC}$
is a pole contribution to the sum rules. The lower bound is found from
convergence of the sum rules. Additionally, quantities extracted from Eqs.\ (%
\ref{eq:Mass}) and (\ref{eq:Coupling}) should be as stable as possible
against variations of $M^{2}$. The continuum threshold parameter $s_{0}$
divides the ground-state contribution and effects of higher resonances and
continuum states. Therefore, $s_{0}$ should be below the first excited state
of the tetraquark $\widetilde{T}_{bc}^{0}$ and obey $\sqrt{s_{0}}-\widetilde{%
m}$ $\approx (400-600)~\mathrm{MeV}$, which can be considered as a reasonable
restriction for heavy tetraquarks.

Performed numerical analyses demonstrate that regions
\begin{equation}
M^{2}\in \lbrack 5.5,7]\ \mathrm{GeV}^{2},\ s_{0}\in \lbrack 58,60]\ \mathrm{%
GeV}^{2},  \label{eq:Wind}
\end{equation}%
satisfy all aforementioned constraints on $M^{2}$ and $s_{0}$. Indeed, at $%
M^{2}=7~\mathrm{GeV}^{2}$ the pole contribution is $37\%$, whereas at $%
M^{2}=5.5~\mathrm{GeV}^{2}$ amounts to $79\%$ of the whole result. These
values of $M^{2}$ fix the boundaries of the region in which the Borel
parameter can be varied. At the minimum of $M^{2}=5.5~\mathrm{GeV}^{2}$
contributions of last three terms to $\Pi (M^{2},s_{0})$ do not exceed $1\%$
of its value.

\begin{widetext}

\begin{figure}[h!]
\begin{center}
\includegraphics[totalheight=6cm,width=8cm]{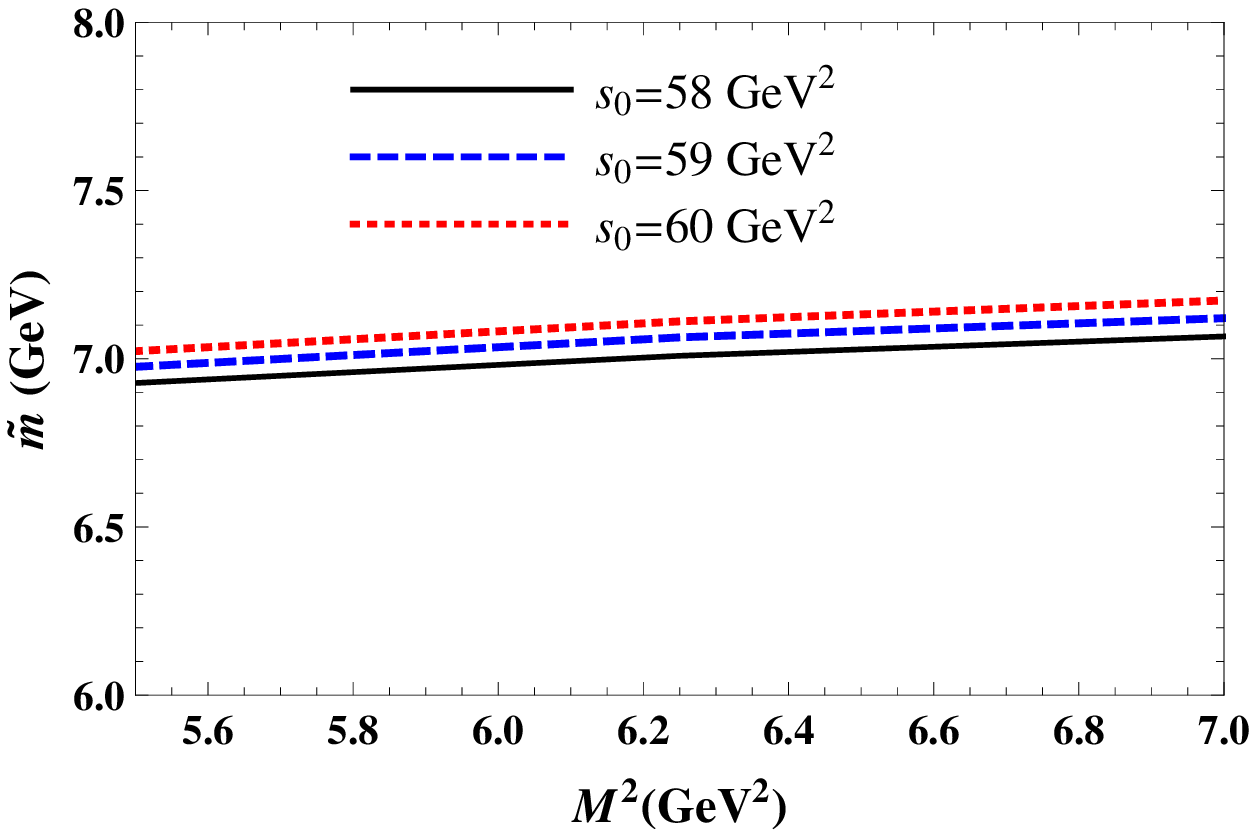}
\includegraphics[totalheight=6cm,width=8cm]{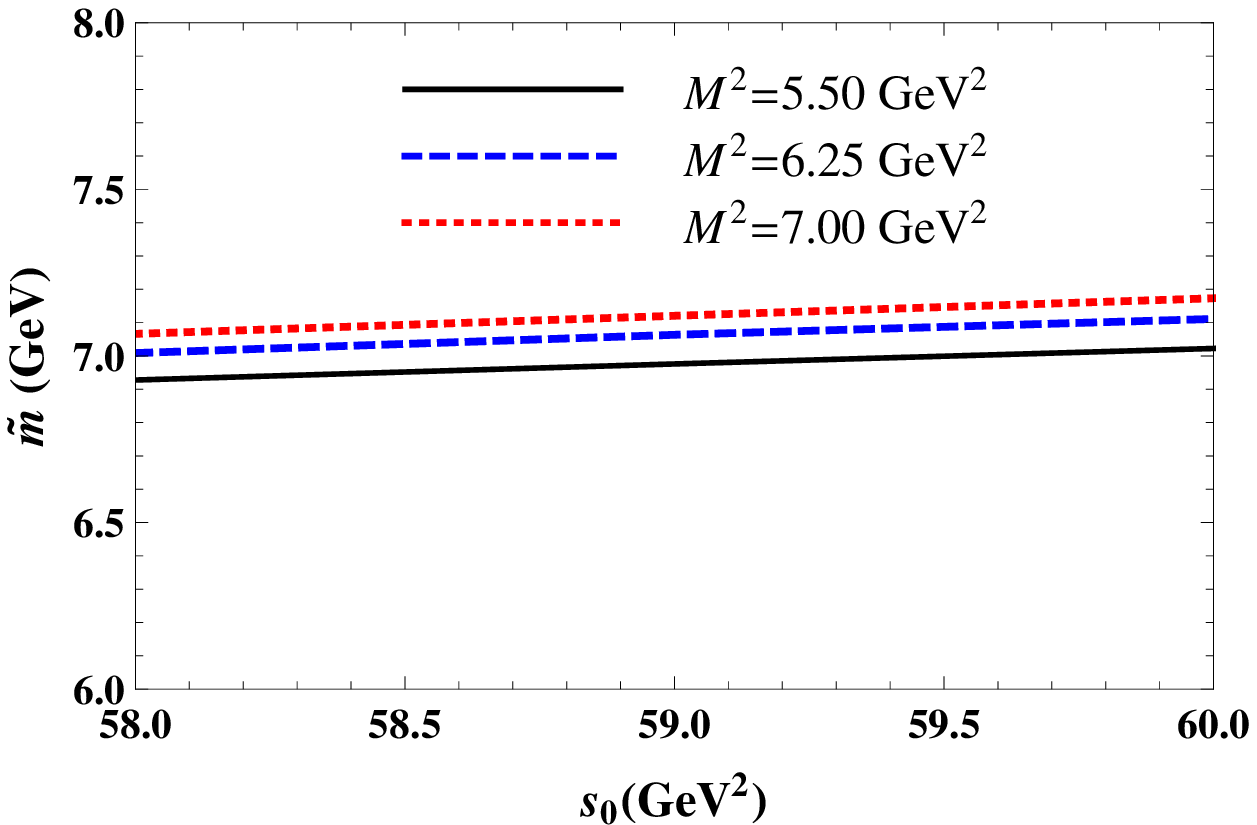}
\end{center}
\caption{ The mass of the tetraquark $\widetilde{T}_{bc}^{0}$ as a function of the Borel parameter
$M^2$ at fixed $s_0$ (left panel) and as a function of the continuum threshold
$s_0$ at fixed $M^2$ (right panel).}
\label{fig:Mass}
\end{figure}

\end{widetext}

For $\widetilde{m}$ and $\widetilde{f}$ we find
\begin{eqnarray}
\widetilde{m} &=&(7050\pm 125)~\mathrm{MeV},  \notag \\
\widetilde{f} &=&(8.3\pm 1.3)\times 10^{-3}~\mathrm{GeV}^{4}.
\label{eq:Result1}
\end{eqnarray}%
In Fig.\ \ref{fig:Mass} we plot the sum rule's prediction for $\widetilde{m}$%
, where its dependence on the Borel $M^{2}$ and continuum threshold
parameter $s_{0}$ is seen explicitly. Theoretical errors in the case of $%
\widetilde{m}$ amount to $\pm 1.8\%$, which confirms a nice accuracy of
performed computations. The ambiguities in deriving of the coupling $%
\widetilde{f}$ are equal to $\pm 16\%$ of the central value: they are larger
than that for $\widetilde{m}$, but still within limits accepted in sum rule
computations. Reasons behind of these effects are quite clear. Indeed, the
sum rule for the mass $\widetilde{m}$ is given by the ratio (\ref{eq:Mass})
which smooths the dependence of $\widetilde{m}$ on the parameter $M^{2}$,
whereas the sum rule for $\widetilde{f}$  (\ref{eq:Coupling}) contains only
the correlator $\Pi (M^{2},s_{0})$.

It is interesting to compare the result obtained for the mass of the
axial-vector tetraquark $\widetilde{T}_{bc}^{0}$ with the mass of $T_{bc}^{0}
$. Let us remind that the latter has the same quark content and quantum
numbers, but is composed of the heavy scalar diquark $b^{T}C\gamma _{5}c$
and light axial-vector antidiquark $\overline{u}C\gamma _{\mu }\overline{d}%
^{T}$. This particle has the mass $(7105\pm 155)~\mathrm{MeV}$ and is $%
\Delta m\approx 50~\mathrm{MeV}$ "heavier" than $\widetilde{T}_{bc}^{0}$.
Our estimate for the mass splitting between $\widetilde{T}_{bc}^{0}$ and $%
T_{bc}^{0}$ is obtained using central values of  their masses.
Spectroscopic parameters of these tetraquarks have been computed in the
context of the QCD sum rule method, predictions of which contain theoretical
errors. It is evident that the mass difference $\Delta m$ is smaller than
uncertainties of this analysis and does not allow one to distinguish these
particles from each other reliably. Real exotic mesons may be superpositions
of these tetraquarks.

An important feature of $\widetilde{T}_{bc}^{0}$\ is that it is stable
against strong decays. In fact, the sum rule method predicts not only the
mass of tetraquarks with some accuracy, but also allows one to estimate
theoretical errors of performed computations. The central values of the
masses $7105~\mathrm{MeV}$\ and $7050~\mathrm{MeV}$ for $%
T_{bc}^{0}$ and $\widetilde{T}_{bc}^{0}$ are below the threshold $\approx
7190~\mathrm{MeV}$ for strong decays to a pair of conventional mesons $%
B^{\ast -}D^{+}$ and $\overline{B}^{\ast 0}D^{0}$. But the maximal
prediction for the mass of $T_{bc}^{0}$ equals to $7260~\mathrm{MeV}$
and exceeds this threshold. Stated differently, the sum rule results do not
exclude strong decays of $T_{bc}^{0}$\ to conventional mesons \cite%
{Agaev:2019kkz}.  In the case of $\widetilde{T}_{bc}^{0}$, we get\ for the
upper bound of the mass $7175~\mathrm{MeV}$ which is still below $7190~\mathrm{MeV}$, and hence $\widetilde{T}_{bc}^{0}$ is strong-interaction
stable particle.

The first excited state of the tetraquark $\widetilde{T}_{bc}^{0}$\ has
presumably a mass $\sqrt{s_{0}}$ $\approx \widetilde{m}+(400-600)~\mathrm{MeV}$
which makes it unstable against decays not only to aforementioned mesons $%
B^{\ast -}D^{+}$ and $\overline{B}^{\ast 0}D^{0}$, but also to pairs $%
B^{-}D^{\ast +}$ and $\overline{B}^{0}D^{\ast 0}$. These two-hadron
reducible terms are common problem when studying multiquark systems in the
framework of the QCD sum rule approach \cite{Kondo:2004cr,Lee:2004xk}. In
the case of tetraquarks two-meson contributions generate the finite width $%
\Gamma (p^{2})$ of these particles and lead to rescaling of their coupling
constants. Nevertheless these effects are small and, as usual, can be 
neglected \cite{Wang:2015nwa,Agaev:2018vag,Sundu:2018nxt}.


\section{Semileptonic decay $T_{bb}^{-} \rightarrow \widetilde{T}_{bc}^{0}l%
\overline{\protect\nu }_{l}$}

\label{sec:WFF}

Weak decays of $T_{bb}^{-}$ can be triggered by subprocesses $b\rightarrow
W^{-}c$ and $b\rightarrow W^{-}u$. The processes generated by the transition
$b\rightarrow W^{-}c$ are dominant decay modes of $T_{bb}^{-}$. The reason
is that the subprocess $b\rightarrow W^{-}u$ leads to decays suppressed
relative to dominant ones by a factor $|V_{bu}|^{2}/|V_{bc}|^{2}$ $\simeq
0.01$, where $V_{q_{1}q_{2}}$ are the Cabibbo-Khobayasi-Maskawa (CKM) matrix
elements. In the present work, we consider only dominant weak decays of $%
T_{bb}^{-}$\ to axial-vector tetraquark $[bc][\overline{u}\overline{d}]$,
which, in general, may contain components $T_{bc}^{0}$\ and $\widetilde{T}%
_{bc}^{0}$. Our analysis proves that the matrix element for the weak
transition $T_{bb}^{-}\rightarrow T_{bc}^{0}$\ vanishes and does not
contribute to relevant decay processes. This fact maybe is connected with
spin flips in initial diquark and antidiquark necessary to form the final
state $T_{bc}^{0}.$ Therefore we restrict ourselves by exploring the weak
transformations $T_{bb}^{-}\rightarrow \widetilde{T}_{bc}^{0}$, which
consist of the semileptonic $T_{bb}^{-}\rightarrow \widetilde{T}_{bc}^{0}l%
\overline{\nu }_{l}$\ and nonleptonic $T_{bb}^{-}$\ $\rightarrow \widetilde{T%
}_{bc}^{0}M$\ processes.

Here, we concentrate on semileptonic decays $T_{bb}^{-}\rightarrow
\widetilde{T}_{bc}^{0}l\overline{\nu }_{l}$ of the tetraquark $T_{bb}^{-}$.
The analysis carried out in the section \ref{sec:Mass} has allowed us to
calculate the spectroscopic parameters of the tetraquark $\widetilde{T}%
_{bc}^{0}$, which are input information to investigate decays of the initial
particle $T_{bb}^{-}$. A big mass gap between the initial and final-state
tetraquarks makes kinematically possible semileptonoc decays with all lepton
spices $l=e,\mu $, and $\tau $.

The effective Hamiltonian to study processes $b\rightarrow W^{-}c$ \ at the
tree-level is determined by the expression
\begin{equation}
\mathcal{H}^{\mathrm{eff}}=\frac{G_{F}}{\sqrt{2}}V_{bc}\overline{c}\gamma
_{\alpha }(1-\gamma _{5})b\overline{l}\gamma ^{\alpha }(1-\gamma _{5})\nu
_{l},  \label{eq:EffecH}
\end{equation}%
where $G_{F}$ and $V_{bc}$ are the Fermi coupling constant and CKM matrix
element, respectively. The matrix element of $\mathcal{H}^{\mathrm{eff}}$
placed between the initial and final tetraquarks contains the leptonic and
hadronic factors%
\begin{equation}
\langle \widetilde{T}_{bc}^{0}(p^{\prime })|\mathcal{H}^{\mathrm{eff}%
}|T_{bb}^{-}(p)\rangle =L^{\alpha }H_{\alpha }.
\end{equation}%
We are interested in calculation of $H_{\alpha }$, because the leptonic part
of the matrix element $L^{\alpha }$ is universal for all semileptonic decays
and does not contain information on tetraquarks. Then, $H_{\alpha }$ is
nothing more than the matrix element of the current
\begin{equation}
J_{\alpha }^{\mathrm{tr}}=\overline{c}\gamma _{\alpha }(1-\gamma _{5})b,
\label{eq:TrCurr}
\end{equation}%
sandwiched between the initial and final particles. It can be modeled in
terms of the form factors $G_{i}(q^{2})$ which parametrize the long-distance
dynamics of the weak transition
\begin{eqnarray}
&&\langle \widetilde{T}_{bc}^{0}(p^{\prime },\epsilon ^{\prime })|J_{\alpha
}^{\mathrm{tr}}|T_{bb}^{-}(p,\epsilon )\rangle =\epsilon ^{\mu }\epsilon
^{\prime \nu }\left[ G_{1}(q^{2})g_{\mu \nu }P_{\alpha }\right.  \notag \\
&&\left. +G_{2}(q^{2})\left( q_{\mu }g_{\alpha \nu }-q_{\nu }g_{\alpha \mu
}\right) -\frac{G_{3}(q^{2})}{2m^{2}}q_{\mu }q_{\nu }P_{\alpha }\right]
\notag \\
&&+G_{4}(q^{2})\varepsilon _{\alpha \mu \rho \nu }\epsilon ^{\mu }\epsilon
^{\prime \rho }P^{\nu },  \label{eq:Vertex}
\end{eqnarray}%
where $(p,\epsilon )$ and $(p^{\prime },\epsilon ^{\prime })$ \ are momenta
and polarization vectors of $T_{bb}^{-}$ and $\widetilde{T}_{bc}^{0}$,
respectively. Here, we also use $P=p+p^{\prime }$ and $q=p-p^{\prime }$. The
momentum transfer in the weak process $q^{2}$ changes within the limits $%
m_{l}^{2}\leq q^{2}\leq (m-\widetilde{m})^{2},$ where $m_{l}$ is the mass of
the lepton $l$.

The weak form factors $G_{i}(q^{2})$ are key ingredients of our
investigations. They should be determined from the QCD three-point sum
rules, which can be derived using the correlation function
\begin{eqnarray}
&&\Pi _{\mu \alpha \nu }(p,p^{\prime })=i^{2}\int d^{4}xd^{4}ye^{i(p^{\prime
}y-px)}  \notag \\
&&\times \langle 0|\mathcal{T}\{\widetilde{J}_{\nu }(y)J_{\alpha }^{\mathrm{%
tr}}(0)J_{\mu }^{^{\dagger }}(x)\}|0\rangle.  \label{eq:CF}
\end{eqnarray}

The standard methods of the sum rule analysis require calculation of the
correlation function $\Pi _{\mu \alpha \nu }(p,p^{\prime })$ using the
physical parameters of the tetraquarks and, by this way, to find the
physical side of the sum rules. At the next phase of studies, one has to
determine $\Pi _{\mu \alpha \nu }(p,p^{\prime })$ by employing the quark
propagators, and express $\Pi _{\mu \alpha \nu }^{\mathrm{OPE}}(p,p^{\prime
})$ in terms of quark, gluon and mixed vacuum condensates. By equating
obtained results and using the assumption about the quark-hadron duality, it
is possible to derive sum rules and compute the form factors of interest.

The physical side of the sum rules $\Pi _{\mu \alpha \nu }^{\mathrm{Phys}%
}(p,p^{\prime })$ can be written down in the following form
\begin{eqnarray}
&&\Pi _{\mu \alpha \nu }^{\mathrm{Phys}}(p,p^{\prime })=\frac{\langle 0|%
\widetilde{J}_{\nu }|\widetilde{T}_{bc}^{0}(p^{\prime },\epsilon ^{\prime
})\rangle \langle \widetilde{T}_{bc}^{0}(p^{\prime },\epsilon ^{\prime
})|J_{\alpha }^{\mathrm{tr}}|T_{bb}^{-}(p,\epsilon )\rangle }{%
(p^{2}-m^{2})(p^{\prime 2}-\widetilde{m}^{2})}  \notag \\
&&\times \langle T_{bb}^{-}(p,\epsilon )|J_{\mu }^{^{\dagger }}|0\rangle
+\cdots,  \label{eq:Phys1}
\end{eqnarray}%
where the contribution of the ground-state particles is shown explicitly,
whereas other terms are denoted by dots.

Calculation of $\Pi _{\mu \alpha \nu }^{\mathrm{Phys}}(p,p^{\prime })$ can
be finished by taking into account Eq.\ (\ref{eq:MElem1}), the explicit
expression of the matrix element $\langle \widetilde{T}_{bc}^{0}(p^{\prime
},\epsilon ^{\prime })|J_{\alpha }^{\mathrm{tr}}|T_{bb}^{-}(p,\epsilon
)\rangle $, and the formula
\begin{equation}
\langle T_{bb}^{-}(p,\epsilon )|J_{\mu }^{^{\dagger }}|0\rangle =fm\epsilon
_{\mu }^{\ast },  \label{eq:MElem2}
\end{equation}%
where $f$ is the coupling of the state $T_{bb}^{-}$. Having substituted the
relevant matrix elements into Eq.\ (\ref{eq:Phys1}), we find the final
expression for $\Pi _{\mu \alpha \nu }^{\mathrm{Phys}}(p,p^{\prime },q^{2})$
\begin{eqnarray}
&&\Pi _{\mu \alpha \nu }^{\mathrm{Phys}}(p,p^{\prime })=\frac{fm\widetilde{f}%
\widetilde{m}}{(p^{2}-m^{2})(p^{\prime 2}-\widetilde{m}^{2})}\left\{
G_{1}(q^{2})p_{\alpha }g_{\mu \nu }\right.  \notag \\
&&+G_{2}(q^{2})\left[ 1-\frac{m^{2}-\widetilde{m}^{2}+q^{2}}{2m^{2}}\right]
p_{\mu }g_{\alpha \nu }  \notag \\
&&\left. -\frac{G_{3}(q^{2})}{2m^{2}}p_{\alpha }p_{\nu }p_{\mu }^{\prime
}+G_{4}(q^{2})\varepsilon _{\theta \alpha \mu \nu }p_{\theta }\right\}
+\cdots.  \label{eq:Phys2}
\end{eqnarray}%
The dots in $\Pi _{\mu \alpha \nu }^{\mathrm{Phys}}(p,p^{\prime })$ stand
for not only effects due to excited and continuum states, but also for
contributions of structures which will not be used in following analysis.

The QCD side of the sum rules can be derived from Eq.\ (\ref{eq:CF}) by
using the interpolating currents and contracting relevant quark fields. The
result of these computations is given by the following formula
\begin{eqnarray}
&&\Pi _{\mu \alpha \nu }^{\mathrm{OPE}}(p,p^{\prime })=i^{2}\int
d^{4}xd^{4}ye^{i(p^{\prime }y-px)}\left\{ \left[ \mathrm{Tr}\left[ \gamma
_{5}\widetilde{S}_{d}^{b^{\prime }b}(x-y)\right. \right. \right.  \notag \\
&&\left. \left. \times \gamma _{5}S_{u}^{a^{\prime }a}(x-y)\right] -\mathrm{%
Tr}\left[ \gamma _{5}\widetilde{S}_{d}^{b^{\prime }a}(x-y)\gamma
_{5}S_{u}^{a^{\prime }b}(x-y)\right] \right]  \notag \\
&&\times \left[ \mathrm{Tr}\left[ \gamma _{\nu }\widetilde{S}%
_{b}^{ia^{\prime }}(-x)(1-\gamma _{5})\gamma _{\alpha }\widetilde{S}%
_{c}^{bi}(y)\gamma _{\mu }S_{b}^{ab^{\prime }}(y-x)\right] \right.  \notag \\
&&\left. \left. -\mathrm{Tr}\left[ \gamma _{\nu }\widetilde{S}%
_{b}^{aa^{\prime }}(y-x)\gamma _{\mu }S_{c}^{bi}(y)\gamma _{\alpha
}(1-\gamma _{5})S_{b}^{ib^{\prime }}(-x)\right] \right] \right\}.  \notag \\
&&
\end{eqnarray}

The correlation function $\Pi _{\mu \alpha \nu }^{\mathrm{OPE}}(p,p^{\prime
})$ contains the same Lorentz structures as its counterpart $\Pi _{\mu
\alpha \nu }^{\mathrm{Phys}}(p,p^{\prime })$. We use corresponding invariant
amplitudes to obtain the required sum rules for the form factors $%
G_{i}(q^{2})$. But before this final operation, we make double Borel
transformation over variables $p^{2}$ and $p^{\prime 2}$ to suppress
contributions of the higher excited and continuum states, and perform
continuum subtraction. These rather routine manipulations give the sum rules
for the form factors $G_{i}(q^{2})$. For $G_{i}(q^{2}),$ $i=1$ and $4$ we
get the similar sum rules%
\begin{eqnarray}
&&G_{i}(M^{2},s_{0},q^{2})=\frac{1}{fm\widetilde{f}\widetilde{m}}%
\int_{4m_{b}^{2}}^{s_{0}}ds\int_{\mathcal{M}^{2}}^{s_{0}^{\prime
}}ds^{\prime }  \notag \\
&&\times \rho _{i}(s,s^{\prime },q^{2})e^{(m^{2}-s)/M_{1}^{2}}e^{(\widetilde{%
m}^{2}-s^{\prime })/M_{2}^{2}},  \label{eq:FF1}
\end{eqnarray}%
where $M_{1}^{2},\ M_{2}^{2}$ and $s_{0},\ s_{0}^{\prime }$ are the Borel
and continuum threshold parameters, respectively. The pair of parameters $%
(M_{1}^{2},s_{0})$ corresponds to a channel of the initial tetraquark $%
T_{bb}^{-}$, whereas $(M_{2}^{2},s_{0}^{\prime })$ describe the final-state
particle $\widetilde{T}_{bc}^{0}$. The remaining two sum rules read:%
\begin{eqnarray}
&&G_{2}(M^{2},s_{0},q^{2})=\frac{2m}{\widetilde{f}\widetilde{m}f(m^{2}+%
\widetilde{m}^{2}-q^{2})}  \notag \\
&&\times \int_{4m_{b}^{2}}^{s_{0}}ds\int_{\mathcal{M}^{2}}^{s_{0}^{\prime
}}ds^{\prime }\rho _{2}(s,s^{\prime },q^{2})e^{(m^{2}-s)/M_{1}^{2}}e^{(%
\widetilde{m}^{2}-s^{\prime })/M_{2}^{2}},  \notag \\
&&{}  \label{eq:FF2}
\end{eqnarray}%
and
\begin{eqnarray}
&&G_{3}(M^{2},s_{0},q^{2})=-\frac{2m}{\widetilde{f}\widetilde{m}f}%
\int_{4m_{b}^{2}}^{s_{0}}ds\int_{\mathcal{M}^{2}}^{s_{0}^{\prime
}}ds^{\prime }  \notag \\
&&\times \rho _{3}(s,s^{\prime },q^{2})e^{(m^{2}-s)/M_{1}^{2}}e^{(\widetilde{%
m}^{2}-s^{\prime })/M_{2}^{2}}.  \label{eq:FF3}
\end{eqnarray}%
As is seen the sum rules are written down using the spectral densities $\rho
_{i}(s,s^{\prime },q^{2})$ which are proportional to the imaginary part of
the corresponding invariant amplitudes in $\Pi _{\mu \alpha \nu }^{\mathrm{%
OPE}}(p,p^{\prime })$. All of them contain both the perturbative and
nonperturbative contributions and are calculated with dimension-5 accuracy.
Explicit expressions of $\rho _{i}(s,s^{\prime },q^{2})$ are cumbersome,
therefore we refrain from providing them here.

The differential rate of the semileptonic decay $T_{bb}^{-} \rightarrow
\widetilde{T}_{bc}^{0}l\overline{\nu }_{l}$ is determined by the weak form
factors $G_{i}(q^{2})$ and is given by the expression \cite{Sundu:2018uyi}
\begin{eqnarray}
&&\frac{d\Gamma }{dq^{2}}=\frac{G_{F}^{2}|V_{cb}|^{2}}{3\cdot 2^{9}\pi
^{3}m^{3}}\left( \frac{q^{2}-m_{l}^{2}}{q^{2}}\right) \lambda \left( m^{2},%
\widetilde{m}^{2},q^{2}\right)  \notag \\
&&\times \left[ \sum_{i=1}^{i=4}G_{i}^{2}(q^{2})\mathcal{A}%
_{i}(q^{2})+G_{1}(q^{2})G_{2}(q^{2})\mathcal{A}_{12}(q^{2})\right.  \notag \\
&&\left. +G_{1}(q^{2})G_{3}(q^{2})\mathcal{A}%
_{13}(q^{2})+G_{2}(q^{2})G_{3}(q^{2})\mathcal{A}_{23}(q^{2})\right. \bigg],
\notag \\
&&  \label{eq:DifW}
\end{eqnarray}%
where
\begin{eqnarray}
&&\lambda \left( m^{2},\widetilde{m}^{2},q^{2}\right) =\left[ m^{4}+%
\widetilde{m}^{4}+q^{4}\right.  \notag \\
&&\left. -2(m^{2}\widetilde{m}^{2}+m^{2}q^{2}+\widetilde{m}^{2}q^{2})\right]
^{1/2}.
\end{eqnarray}%
The decay rate $d\Gamma /dq^{2}$ depends also on functions $\mathcal{A}%
_{i}(q^{2})$ and $\mathcal{A}_{ij}(q^{2})$ which can be found in Ref.\ \cite%
{Sundu:2018uyi}.

Sum rules for $G_{i}(q^{2})$ are necessary to find corresponding fit
functions $\mathcal{G}_{i}(q^{2})$ and calculate the width of the
semileptonic decays. Technical details of numerical computations to extract
weak form factors are well known. In fact, Eqs.\ (\ref{eq:FF1}), (\ref%
{eq:FF2}) and (\ref{eq:FF3}) through the spectral densities $\rho
_{i}(s,s^{\prime },q^{2})$ depend on the quark, gluon and mixing
condensates, as well as masses of the $c$ and $b$-quarks: these parameters
have been specified in the previous section. Besides, the sum rules contain
also masses and couplings of the tetraquarks $T_{bb}^{-}$ and $\widetilde{T}%
_{bc}^{0}$. The mass and coupling of $T_{bb}^{-}$ were evaluated in Ref.\
\cite{Agaev:2018khe}%
\begin{eqnarray}
m &=&(10035\pm 260)~\mathrm{MeV},  \notag \\
f &=&(1.38\pm 0.27)\times 10^{-2}~\mathrm{GeV}^{4}.  \label{eq:Result2}
\end{eqnarray}%
The spectroscopic parameters of the tetraquark $\widetilde{T}_{bc}^{0}$ have
been found in the present work. We need also to fix Borel and continuum
threshold parameters to carry out numerical analysis. In the intial particle
channel $(M_{1}^{2},s_{0})$ are chosen as in Ref.\ \cite{Agaev:2018khe}, in
which the mass and coupling of $T_{bb}^{-}$ were calculated%
\begin{equation}
M^{2}\in \lbrack 9,13]\ \mathrm{GeV}^{2},\ s_{0}\in \lbrack 115,120]\
\mathrm{GeV}^{2}.  \label{eq:Wind2}
\end{equation}%
For the next pair $(M_{2}^{2},s_{0}^{\prime })$ we use parameters from Eq.\ (%
\ref{eq:Wind}).

The sum rules give reliable results for $G_{i}(q^{2})$ in the region $%
m_{l}^{2}\leq q^{2}\leq 7~\mathrm{GeV}^{2}$, which is not enough to find the
partial width of the process $T_{bb}^{-}\rightarrow \widetilde{T}_{bc}^{0}l%
\overline{\nu }_{l}$ under consideration. To calculate the width of the
semileptonic decay $d\Gamma /dq^{2}$ must be integrated over $q^{2}$ in the
boundaries $m_{l}^{2}\leq q^{2}\leq (m-\widetilde{m})^{2}$, i.e., in the
limits $m_{l}^{2}\leq q^{2}\leq 8.9~\mathrm{GeV}^{2}$. But this region is
wider than the one where the sum rules lead to strong results. This problem
can be solved by introducing extrapolating (fit) functions $\mathcal{G}%
_{i}(q^{2})$. At the momentum transfers $q^{2}$ accessible for the sum rule
computations they must coincide with $G_{i}(q^{2})$, but have analytic forms
suitable to carry out integrations over $q^{2}$.
\begin{figure}[h]
\includegraphics[width=8.8cm]{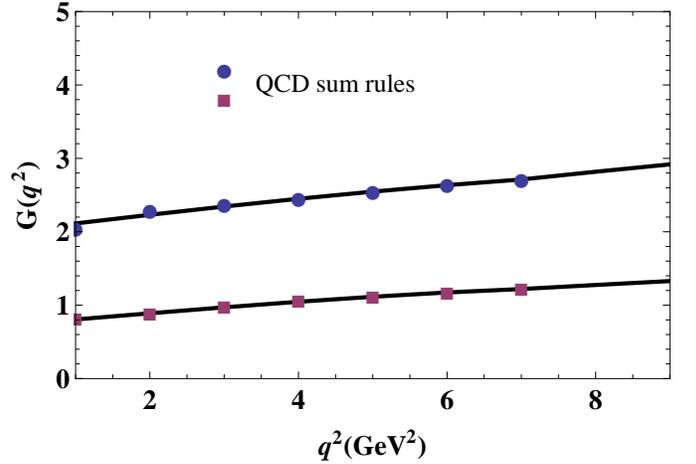}
\caption{Sum rule predictions for the weak form factors $G_{1}(q^{2})$ (the
upper blue circles) and $|G_{2}(q^{2})|$ (the lower red squares). The lines
denote the fit functions $\mathcal{G}_{1}(q^{2})$ and $|\mathcal{G}%
_{2}(q^{2})|$, respectively.}
\label{fig:GFF}
\end{figure}

To this end, we use the functions
\begin{equation}
\mathcal{G}_{i}(q^{2})=\mathcal{G}_{0}^{i}\exp \left[ g_{1}^{i}\frac{q^{2}}{%
m^{2}}+g_{2}^{i}\left( \frac{q^{2}}{m^{2}}\right) ^{2}\right] ,
\label{eq:FFunctions}
\end{equation}%
where parameters $\mathcal{G}_{0}^{i},~g_{1}^{i}$, and $g_{2}^{i}$ should be
fitted to satisfy sum rules' predictions. The parameters of the functions $%
\mathcal{G}_{i}(q^{2})$ extracted from numerical analysis are collected in
Table\ \ref{tab:ExtraF}. As an example, the functions $\mathcal{G}%
_{1}(q^{2}) $ and $|\mathcal{G}_{2}(q^{2})|$ are depicted in Fig.\ \ref%
{fig:GFF}, in which we see a quite nice agreement between the sum rule
predictions and fit functions.

To calculate the partial widths of the semileptonic decays, apart from the
form factors, we also use the parameters
\begin{eqnarray}
G_{F} &=&1.16637\times 10^{-5}~\mathrm{GeV}^{-2},  \notag \\
|V_{bc}| &=&(42.2\pm 0.08)\times 10^{-3}.
\end{eqnarray}

\begin{table}[tbp]
\begin{tabular}{|c|c|c|c|}
\hline\hline
$\mathcal{G}_{i}(q^{2})$ & $\mathcal{G}_{0}^{i}$ & $g_{1}^{i}$ & $g_{2}^{i}$
\\ \hline
$\mathcal{G}_{1}(q^{2})$ & $2.13$ & $3.28$ & $-4.33$ \\
$\mathcal{G}_{2}(q^{2})$ & $-0.72$ & $6.26$ & $-17.13$ \\
$\mathcal{G}_{3}(q^{2})$ & $334.57$ & $3.31$ & $-5.71$ \\
$\mathcal{G}_{4}(q^{2})$ & $-1.33$ & $3.36$ & $0.35$ \\ \hline\hline
\end{tabular}%
\caption{Parameters of the extrapolating functions $\mathcal{G}_{i}(q^{2})$.}
\label{tab:ExtraF}
\end{table}

Our results for the partial widths of the semileptonic decay channels are
presented below
\begin{eqnarray}
&&\Gamma (T_{bb}^{-}\rightarrow \widetilde{T}_{bc}^{0}e^{-}\overline{\nu }%
_{e})=(2.02\pm 0.39)\times 10^{-9}~\mathrm{MeV},  \notag \\
&&\Gamma (T_{bb}^{-}\rightarrow \widetilde{T}_{bc}^{0}\mu ^{-}\overline{\nu }%
_{\mu })=(1.96\pm 0.37)\times 10^{-9}~\mathrm{MeV},  \notag \\
&&\Gamma (T_{bb}^{-}\rightarrow \widetilde{T}_{bc}^{0}\tau ^{-}\overline{\nu
}_{\tau })=(1.03\pm 0.19)\times 10^{-10}~\mathrm{MeV}.  \notag \\
&&  \label{eq:Results}
\end{eqnarray}%
Obtained information on partial widths of semileptonic channels can be used
to improve predictions for the full width and lifetime of the exotic meson $%
T_{bb}^{-}$.


\section{ Nonleptonic processes $T_{bb}^{-} \rightarrow \widetilde{T}%
_{bc}^{0}M$}

\label{sec:NLDecays}

Dominant nonleptonic decays of $T_{bb}^{-}$ are triggered by the subprocess $%
b\rightarrow W^{-}c$, whereas the transition $b\rightarrow W^{-}u$ leads to
decays suppressed relative to main ones, as it has been explained in the
previous section. Therefore, in this section we consider nonleptonic decays $%
T_{bb}^{-} \rightarrow \widetilde{T}_{bc}^{0}M$ of the tetraquark $%
T_{bb}^{-} $. In these processes $M$ is one of the pseudoscalar mesons $\pi
^{-}$, $K^{-}$, $\ D^{-}$, and $\ D_{s}^{-}$. At the final state of the
process they appear due to decays of $W^{-}$ to pairs of quark-antiquark $d%
\overline{u}$, $s\overline{u}$, $d\overline{c}$, and $s\overline{c}$,
respectively. The masses and decay constants of the mesons $\pi ^{-}$, $%
K^{-} $, $D^{-}$, and $D_{s}^{-}$ are presented in Table\ \ref{tab:MesonPar}%
. It is not difficult to see, that the mass of $T_{bb}^{-}$ obeys a
requirement $m>\widetilde{m}+m_{M}$ for all of these mesons, and these
decays are kinematically allowed processes.

We describe production of mesons $M$ by employing the effective Hamiltonian,
and introducing relevant effective weak vertices. To study the nonleptonic
decays $T_{bb}^{-} \rightarrow \widetilde{T}_{bc}^{0}M$, we use also the QCD
factorization approach. This method is fruitful for studying of ordinary
mesons' nonleptonic decays \cite{Beneke:1999br,Beneke:2000ry}, but can be
applied to investigate weak decays of tetraquarks as well \cite%
{Agaev:2019lwh,Sundu:2019feu}.

We present in a detailed form the decay $T_{bb}^{-} \rightarrow \widetilde{T}%
_{bc}^{0}\pi ^{-}$, and write down final results for other channels. The
effective Hamiltonian $\widetilde{\mathcal{H}}^{\mathrm{eff}}$ for this
decay at the tree-level is given by the expression
\begin{equation}
\widetilde{\mathcal{H}}^{\mathrm{eff}}=\frac{G_{F}}{\sqrt{2}}%
V_{bc}V_{ud}^{\ast }\left[ c_{1}(\mu )Q_{1}+c_{2}(\mu )Q_{2}\right] ,
\label{eq:EffHam}
\end{equation}%
where%
\begin{eqnarray}
Q_{1} &=&\left( \overline{d}_{i}u_{i}\right) _{\mathrm{V-A}}\left( \overline{%
c}_{j}b_{j}\right) _{\mathrm{V-A}},  \notag \\
Q_{2} &=&\left( \overline{d}_{i}u_{j}\right) _{\mathrm{V-A}}\left( \overline{%
c}_{j}b_{i}\right) _{\mathrm{V-A}},  \label{eq:Operators}
\end{eqnarray}%
and $i$ , $j$ are the color indices. In Eq.\ (\ref{eq:EffHam}) the
abbreviation $\left( \overline{q}_{1}q_{2}\right) _{\mathrm{V-A}}$ means
\begin{equation}
\left( \overline{q}_{1}q_{2}\right) _{\mathrm{V-A}}=\overline{q}_{1}\gamma
_{\mu }(1-\gamma _{5})q_{2}.  \label{eq:Not}
\end{equation}%
Let us note that, we do not include into Eq.\ (\ref{eq:EffHam})
current-current operators appearing due the QCD penguin and
electroweak-penguin diagrams. The short-distance Wilson coefficients $%
c_{1}(\mu )$ and $c_{2}(\mu )$ are given at the factorization scale $\mu $.

The amplitude of the decay $T_{bb}^{-} \rightarrow \widetilde{T}_{bc}^{0}\pi
^{-}$ is determined by the following expression%
\begin{eqnarray}
\mathcal{A} &=&\frac{G_{F}}{\sqrt{2}}V_{bc}V_{ud}^{\ast }a_{1}(\mu )\langle
\pi ^{-}(q)|\left( \overline{d}_{i}u_{i}\right) _{\mathrm{V-A}}|0\rangle
\notag \\
&&\times \langle \widetilde{T}_{bc}^{0}(p^{\prime })|\left( \overline{c}%
_{j}b_{j}\right) _{\mathrm{V-A}}|T_{bb}^{-}(p)\rangle,  \label{eq:Amplitude}
\end{eqnarray}%
where
\begin{equation}
a_{1}(\mu )=c_{1}(\mu )+\frac{1}{N_{c}}c_{2}(\mu ),
\end{equation}%
with $N_{c}=3$ being the number of quark colors.

The matrix element $\langle \widetilde{T}_{bc}^{0}(p^{\prime })|\left(
\overline{c}_{j}b_{j}\right) _{\mathrm{V-A}}|T_{bb}^{-}(p)\rangle $ in terms
of the weak form factors is given by Eq.\ (\ref{eq:Vertex}). The matrix
element $\langle \pi ^{-}(q)|\left( \overline{d}_{i}u_{i}\right) _{\mathrm{%
V-A}}|0\rangle $ in $\mathcal{A}$ can be written down in the following form
\begin{equation}
\langle \pi ^{-}(q)|\left( \overline{d}_{i}u_{i}\right) _{\mathrm{V-A}%
}|0\rangle =if_{\pi }q_{\mu },  \label{eq:ME4}
\end{equation}%
where $f_{\pi }$ is the decay constant of the pion. Then, the amplitude $%
\mathcal{A}$ of the nonleptonic weak decay is determined by the expression
\begin{eqnarray}
&&\mathcal{A}=i\frac{G_{F}}{\sqrt{2}}f_{\pi }V_{bc}V_{ud}^{\ast }a_{1}(\mu
)\left\{ Pq\left[ G_{1}(q^{2})\epsilon \cdot \epsilon ^{\prime }\right.
\right.  \notag \\
&&\left. \left. -\frac{G_{3}(q^{2})}{2m^{2}}q\cdot \epsilon q\cdot \epsilon
^{\prime }\right] +G_{4}(q^{2})\varepsilon _{\alpha \mu \rho \nu }\epsilon
^{\mu }\epsilon ^{\prime \rho }P^{\nu }q^{\alpha }\right\} .  \notag \\
&&  \label{eq:Amplitude2}
\end{eqnarray}%
For completeness we provide below the partial width of this process%
\begin{equation}
\Gamma (T_{b:\overline{d}}^{-}\rightarrow \widetilde{Z}_{bc}^{0}\pi ^{-})=%
\frac{|\mathcal{A}|^{2}}{48\pi m^{3}}\lambda \left( m^{2},\widetilde{m}%
_{Z}^{2},q^{2}\right) ,  \label{eq:NLDW}
\end{equation}%
where
\begin{eqnarray}
&&|\mathcal{A}|^{2}=\frac{G_{F}^{2}}{2}f_{\pi
}^{2}|V_{bc}|^{2}|V_{ud}|^{2}a_{1}^{2}(\mu )\left\{ G_{1}^{2}(q^{2})\left(
m^{2}-\widetilde{m}^{2}\right) ^{2}\right.  \notag \\
&&\times \frac{\left[ m^{4}+\left( \widetilde{m}^{2}-q^{2}\right)
^{2}-2m^{2}\left( 5\widetilde{m}^{2}-q^{2}\right) \right] }{4m^{2}\widetilde{%
m}^{2}}  \notag \\
&&+G_{3}^{2}(q^{2})\frac{\left( m^{2}-\widetilde{m}^{2}\right) ^{2}}{64m^{6}%
\widetilde{m}^{2}}\left[ m^{4}+\left( \widetilde{m}^{2}-q^{2}\right)
^{2}\right.  \notag \\
&&\left. -2m^{2}\left( \widetilde{m}^{2}+q^{2}\right) \right]
^{2}+2G_{4}^{2}(q^{2})\left[ m^{4}+\left( \widetilde{m}^{2}-q^{2}\right)
^{2}\right.  \notag \\
&&\left. -2m^{2}\left( \widetilde{m}^{2}+q^{2}\right) \right]
+G_{1}(q^{2})G_{3}(q^{2})\frac{\left( m^{2}-\widetilde{m}^{2}\right) ^{2}}{%
8m^{4}\widetilde{m}^{2}}  \notag \\
&&\times \left[ m^{6}+\left( \widetilde{m}^{2}-q^{2}\right) ^{3}-m^{4}\left(
\widetilde{m}^{2}+3q^{2}\right) \right.  \notag \\
&&\left. \left. -m^{2}\left( \widetilde{m}^{4}+2\widetilde{m}%
^{2}q^{2}-3q^{4}\right) \right] \right\} .  \label{eq:A2}
\end{eqnarray}%
In Eqs. (\ref{eq:NLDW}) and (\ref{eq:A2}) the weak form factors $%
G_{i}(q^{2}) $\ and $\lambda \left( m^{2},\widetilde{m}_{Z}^{2},q^{2}\right)
$ are computed at $q^{2}=m_{\pi }^{2}$. The decay modes $T_{bb}^{-}$\ $%
\rightarrow \widetilde{T}_{bc}^{0}K^{-}(D^{-},\ D_{s}^{-})$\ can be studied
in a similar way. For these purposes, in expressions above one should
replace ($m_{\pi },f_{\pi }$) by the masses and decay constants of the
mesons $K^{-}$, $D^{-}$, and $D_{s}^{-}$, make the substitutions $%
V_{ud}\rightarrow V_{us}$, $V_{cd} $, and $V_{cs}$, and fix the form factors
$G_{i}$ and $\lambda $ at $q^{2}=m_{M}^{2}$.

Input parameters required for numerical computations are collected in Table %
\ref{tab:MesonPar}. This table contains the masses and decay constants of
the final state mesons, and relevant CKM matrix elements. The coefficients $%
c_{1}(m_{b})$, and $c_{2}(m_{b})$ with next-to-leading order QCD corrections
are borrowed from Refs.\ \cite{Buras:1992zv,Ciuchini:1993vr,Buchalla:1995vs}
\begin{equation}
c_{1}(m_{b})=1.117,\ c_{2}(m_{b})=-0.257.  \label{eq:WCoeff}
\end{equation}

\begin{table}[tbp]
\begin{tabular}{|c|c|}
\hline\hline
Quantity & Value \\ \hline\hline
$m_{\pi} $ & $139.570~\mathrm{MeV}$ \\
$m_{K}$ & $(493.677\pm 0.016)~\mathrm{MeV}$ \\
$m_{D}$ & $(1869.61 \pm 0.10)~\mathrm{MeV}$ \\
$m_{D_s}$ & $(1968.30\pm 0.11)~\mathrm{MeV}$ \\
$f_{\pi }$ & $131~\mathrm{MeV}$ \\
$f_{K}$ & $(155.72\pm 0.51)~\mathrm{MeV}$ \\
$f_{D}$ & $(203.7 \pm 4.7)~\mathrm{MeV}$ \\
$f_{D_s}$ & $(257.8 \pm 4.1)~\mathrm{MeV}$ \\
$|V_{ud}|$ & $0.97420\pm 0.00021$ \\
$|V_{us}|$ & $0.2243\pm 0.0005$ \\
$|V_{cd}|$ & $0.218\pm 0.004$ \\
$|V_{cs}|$ & $0.997\pm 0.017$ \\ \hline\hline
\end{tabular}%
\caption{Masses and decay constants of the final state pseudoscalar mesons.
The CKM matrix elements are also included. }
\label{tab:MesonPar}
\end{table}

For the decay $T_{bb}^{-}\rightarrow \widetilde{T}_{bc}^{0}\pi ^{-}$
calculations yield
\begin{eqnarray}
&&\Gamma (T_{bb}^{-}\rightarrow \widetilde{T}_{bc}^{0}\pi ^{-})=\left(
5.84\pm 1.11\right) \times 10^{-10}~\mathrm{MeV}.  \notag \\
&&  \label{eq:NLDW1}
\end{eqnarray}%
Partial widths of other nonleptonic decays of the tetraquark $T_{bb}^{-}$
are
\begin{eqnarray}
&&\Gamma (T_{bb}^{-}\rightarrow \widetilde{T}_{bc}^{0}K^{-})=\left( 6.43\pm
1.32\right) \times 10^{-11}~\mathrm{MeV},  \notag \\
&&\Gamma (T_{bb}^{-}\rightarrow \widetilde{T}_{bc}^{0}D^{-})=\left( 3.01\pm
0.64\right) \times 10^{-11}~\mathrm{MeV},  \notag \\
&&\Gamma (T_{bb}^{-}\rightarrow \widetilde{T}_{bc}^{0}D_{s}^{-})=\left(
7.80\pm 1.54\right) \times 10^{-10}~\mathrm{MeV}.  \notag \\
&&  \label{eq:ResultsB}
\end{eqnarray}%
Results of this section are second part of required information.


\section{Discussion and concluding notes}

\label{sec:Disc}

Results of the previous two sections allow us to improve predictions for the
full width and mean lifetime of the tetraquark $T_{bb}^{-}$. The
semileptonic decays of this particle $T_{bb}^{-} \rightarrow Z_{bc}^{0}l%
\overline{\nu }_{l}$ were analyzed in Ref.\ \cite{Agaev:2018khe}. Using
partial widths of these processes and widths of the semileptonic and
nonleptonic decays $T_{bb}^{-} \rightarrow \widetilde{T}_{bc}^{0}l\overline{%
\nu }_{l}$ and $T_{bb}^{-} \rightarrow \widetilde{T}_{bc}^{0}M$, it is not
difficult to evaluate relevant parameters. Thus, for the full width of $%
T_{bb}^{-}$, we get
\begin{equation}
\widetilde{\Gamma }=(7.72\pm 1.23)\times 10^{-8}~\mathrm{MeV}.  \label{eq:FW}
\end{equation}%
The lifetime of $T_{bb}^{-}$ is estimated in the range%
\begin{equation}
\widetilde{\tau }=8.53_{-1.18}^{+1.57}~\mathrm{fs}.  \label{eq:LT}
\end{equation}

Comparing improved estimates for $\widetilde{\Gamma}$ and $\widetilde{\tau}$
with previous results
\begin{eqnarray}
\Gamma &=&(7.17\pm 1.23)\times 10^{-8}~\mathrm{MeV},  \notag \\
\tau &=&9.18_{-1.34}^{+1.90}~\mathrm{fs},  \label{eq:PreviousR}
\end{eqnarray}%
one can see that semileptonic processes $T_{bb}^{-} \rightarrow Z_{bc}^{0}l%
\overline{\nu }_{l}$ are dominant decay channels of the tetraquark $%
T_{bb}^{-}$. In fact, a difference $\Delta =\widetilde{\Gamma }-\Gamma
=0.55\times 10^{-8}~\mathrm{MeV}$ is equal to $8\%$ of $\Gamma $. In other
words, $7$ new decay modes considered in the present work constitute
approximately $8\%$ part of the full width $\Gamma $. The branching ratios
of different weak decay modes of $T_{bb}^{-}$ are presented in Table \ref%
{tab:BR}, excluding two nonleptonic decays $\mathcal{BR}$s of which are
negligible.

\begin{table}[tbp]
\begin{tabular}{|c|c|}
\hline\hline
Channels & $\mathcal{BR}(\%)$ \\ \hline\hline
$T_{bb}^{-} \rightarrow Z_{bc}^{0}e^{-}\overline{\nu }_{e}$ & $34.3$ \\
$T_{bb}^{-} \rightarrow Z_{bc}^{0}\mu ^{-}\overline{\nu }_{\mu}$ & $34.2$ \\
$T_{bb}^{-} \rightarrow Z_{bc}^{0}\tau ^{-}\overline{\nu }_{\tau }$ & $24.4$
\\
$T_{bb}^{-} \rightarrow \widetilde{T}_{bc}^{0}e^{-}\overline{\nu }_{e}$ & $%
2.6$ \\
$T_{bb}^{-} \rightarrow \widetilde{T}_{bc}^{0}\mu ^{-}\overline{\nu }_{\mu}$
& $2.5$ \\
$T_{bb}^{-} \rightarrow \widetilde{T}_{bc}^{0}\tau ^{-}\overline{\nu }_{\tau
}$ & $0.13$ \\
$T_{bb}^{-} \rightarrow \widetilde{T}_{bc}^{0}\pi^{-}$ & $0.76$ \\
$T_{bb}^{-} \rightarrow \widetilde{T}_{bc}^{0}D_{s}^{-}$ & $1.01$ \\
\hline\hline
\end{tabular}%
\caption{Branching ratios of the weak decay channels of the tetraquark $%
T_{bb}^{-}$.}
\label{tab:BR}
\end{table}

We have explored the weak decays of $T_{bb}^{-}$, where the final-state
tetraquark $\widetilde{T}_{bc}^{0}=[bc][\overline{u}\overline{d}]$ has been
considered as an axial-vector particle. By computing partial widths of these
decays, we have calculated the full width $\widetilde{\Gamma }$ and mean
lifetime $\widetilde{\tau }$ of the axial-vector tetraquark $T_{bb}^{-}$ and
improved existing estimations for them. Up to now experimental
collaborations did not discover weak decays of tetraquarks. But some of
active experiments, such as LHC, have a certain potential to observe weak
decay channels of tetraquarks $T_{bb}$ \cite{Ali:2018xfq}: In Ref.\ \cite%
{Ali:2018xfq} the authors demonstrated that during $1-4$ runs of LHC one may
expect $\mathcal{O}(10^{8})$ events with $T_{bb}^{-}$. Such potential will
have also a Tera-$Z$ factory \cite{Ali:2018ifm}. Predictions obtained for
parameters of the tetraquark $T_{bb}^{-}$ may be useful for analysis of
these processes.

\section*{ACKNOWLEDGEMENTS}

The work of K.~A, B.~B., and H.~S was supported in part by TUBITAK under the
grant No: 119F050.

\end{document}